\documentclass[9pt]{article}
\usepackage{spconf,amsmath,graphicx}
\usepackage[backref=false]{hyperref}
\hypersetup{hidelinks} 
\usepackage{float} % for H
\usepackage{multirow} % for table
\usepackage[hang]{subfigure} % for caption linebreak aligning
\usepackage{microtype}
\usepackage{enumitem}
\usepackage{adjustbox}
\usepackage[skip=2pt]{caption}

\usepackage{fancyhdr}[]
\fancypagestyle{firstpage}{
\lhead{}
\pagestyle{fancy}
    \chead{}
    \rhead{}
    \lfoot{}
    \cfoot{\small © 2024 IEEE. Personal use of this material is permitted. Permission from IEEE must be obtained for all other uses, in any current or future media, including reprinting/republishing this material for advertising or promotional purposes, creating new collective works, for resale or redistribution to servers or lists, or reuse of any copyrighted component of this work in other works.}
    \rfoot{}

}
\title{Music source separation based on a lightweight deep learning framework (DTTNET: DUAL-PATH TFC-TDF UNET)}
\name{Junyu Chen$^{\star}$ \qquad Susmitha Vekkot$^{\dagger}$ \qquad Pancham Shukla$^{\star}$}

\address{$^{\star}$ Department of Computing, Imperial College London, UK \\
$^{\dagger}$Department of Electronics \& Communication Engineering, \\ Amrita School of Engineering, Bengaluru, Amrita Vishwa Vidyapeetham, India}

\begin{document}
%\ninept
%
\maketitle

\begin{abstract}
% Music source separation (MSS) aims to separate a specific source from a music mixture using the supervised learning method. Although many methods have already been proposed and achieved an excellent result, the model size is becoming bigger and bigger. In this paper, we propose a novel architecture called called DTNet (Dual-path TFC-TDF UNET). We show that with fewer parameters the model can still achieve a leading performance. And our code is available at xxx.

% Music source separation (MSS) aims to extract a designated source from a composite musical blend using supervised learning methods. Despite the myriad of deep learning methods that have been introduced, yielding remarkable outcomes, there's an observable trend towards the expansion of model sizes. In this paper, we introduce a novel architecture called Dual-path TFC-TDF UNET (DTTNet). We substantiate that even with a 20\% of parameter count of BSRNN (SOTA), the model obtains the 9.95 dB cSDR on vocal track compared to 10.01 dB from BSRNN. Additionally, we conducted pattern specific evaluation and analyzed model generalization ability in more complex scenarios.

\noindent Music source separation (MSS) aims to extract 'vocals', 'drums', 'bass' and 'other' tracks from a piece of mixed music. While deep learning methods have shown impressive results, there is a trend toward larger models. In our paper, we introduce a novel and lightweight architecture called DTTNet\footnote{The code can be accessed from: \url{https://github.com/junyuchen-cjy/DTTNet-Pytorch}.}, which is based on Dual-Path Module and Time-Frequency Convolutions Time-Distributed Fully-connected UNet (TFC-TDF UNet). DTTNet achieves 10.12 dB cSDR on 'vocals' compared to 10.01 dB reported for Bandsplit RNN (BSRNN) but with 86.7\% fewer parameters. We also assess pattern-specific performance and model generalization for intricate audio patterns.

% result
% code
\end{abstract}
% \begin{keywords}
% One, two, three, four, five
% \end{keywords}
%
\begin{figure*}[t]
    \centering
    \includegraphics[width=\linewidth]{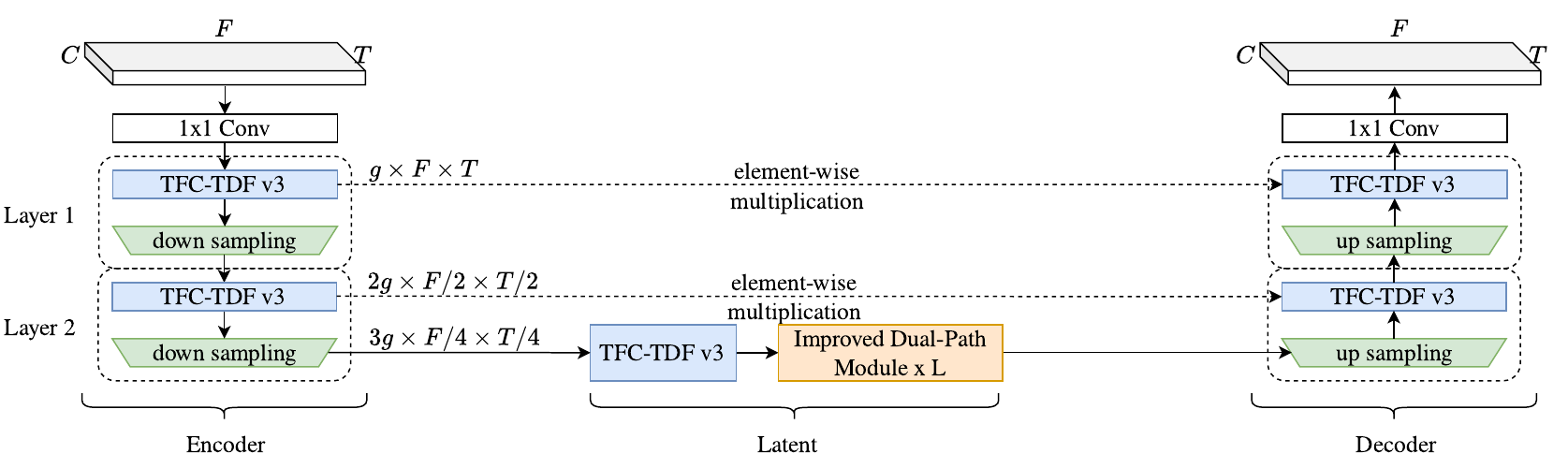}
    \caption{A framework of Dual Path TFC-TDF UNet when layer depth $D = 2$, where $L$ is the number of repeats of Improved Dual-Path Module (IDPM); $C$ as the number of channels of input spectrogram with $g$ being the channel incremental factor; $T$ and $F$ are the time and frequency axes that 2D convolution operates on. }
    \label{fig:arch}
\end{figure*}
\section{Introduction}
\label{sec:intro}
% 介绍问题
% 介绍现有方法
% 指出现有方法的不足
% 提出我们的解决方案
Music source separation (MSS) separates a specific target waveform $s_{target} \in R^{C \times T}$ from a mixture waveform $s_{mix} \in R^{C \times T}$~\cite{SiSEC18}. This problem is similar to the "cocktail party effect"~\cite{cocktail}, where humans can focus on a specific speaker/instrument in a noisy environment. In particular, if the target waveform is vocal, then this subtask is named Singing Voice Separation (SVS)~\cite{choiInvestigatingUNetsVarious2020}. The separated vocals improve the performance of pitch tracking algorithms, which are vital for tasks like pitch correction and speech analysis~\cite{kim_crepe_2018}.

In the realm of deep learning, MSS can be reformulated as a regression problem~\cite{mitsufuji_music_2022}. For the waveform domain models~\cite{stoller_wave-u-net_2018, defossez_music_2021}, the input and output of the neural network are both audio waveforms in $R^{C \times T}$. For the frequency domain models~\cite{stoter19}, the model operates on the spectrogram from Short Time Fourier Transform (STFT). To recover the signal, the Inverse Short Time Fourier Transform (ISTFT) is applied to the predicted spectrogram. The input of these models used to be real-value spetrogram~\cite{ spleeter2020, takahashi_mmdenselstm_2018}. But recent state-of-the-art models~\cite{kimKUIELabMDXNetTwoStreamNeural2021, luo_music_2022, kim_sound_2023} focus on complex domain spectrogram. The studies in~\cite{kong_decoupling_2021} show that complex spectrograms are well suited for improved Source-to-Distortion Ratio (SDR) over real-valued spectrograms.

The current state-of-the-art models for separating the 'vocals' track in MSS problems are Band-split Recurrent Neural Network (BSRNN)~\cite{luo_music_2022} and Time-Frequency Convolutions Time-Distributed Fully-connected UNet (TFC-TDF UNet) v3~\cite{kim_sound_2023}. BSRNN predicts a complex mask on the spectrogram and uses fully connected layers (FC) as well as multilayer perceptron  (MLP) to encode and decode the features. The encoded features are further processed by 12 Dual-Path RNNs to capture the inter and intra dependencies across subbands. However, the FC and MLP layers introduce a large number of redundant parameters and the 12-layer Dual-Path RNNs require increased training time. 
TFC-TDF UNet v3 uses residual convolution blocks. Moreover, TFC-TDF UNet v3 does not introduce explicit time modeling, and therefore the performance gain is less prominent when the parameters of the models are increased drastically.

In this paper, we introduce a novel and lightweight framework called DTTNet, which is based on Dual-Path Module and TFC-TDF UNet v3. The contributions of this work are as follows:
% DTTNet has fewer Dual-Path layers and residual convolution layers. We demonstrate that by incorporating residual convolution encoding and decoding layers, even with 4 layers of Dual-Path module, we can substantially maintain top-tier performance on the 'vocal' and 'other' track while reducing the parameter size. Besides, we enhance the parallelization within the Dual-Path module by splitting the feature dimension and significantly reducing inference time.
% Although Dual-Path modules effectively capture time and frequency dependencies, with the increase of feature dimension, they come with a significant training time overhead. To address this issue, we partition the features and enhance parallelization within the Dual-Path module. 
% Additionally, we demonstrate that by incorporating residual convolution layers, even with 4 layers of Dual-Path module, we can substantially maintain top-tier performance on the 'vocal' and 'other' track while reducing the parameter size. 
% Additionally, the MUSDB18-HQ~\cite{MUSDB18} dataset, although popular for separation studies, doesn't closely resemble modern pop songs. Hence, models trained on it may struggle to generalize. To overcome this, we collected 5 categories of patterns and examined the model's performance in detail.

\begin{figure*}[t]
    \centering
    \subfigure[TFC-TDF v3]{
        \includegraphics[height=0.45\linewidth]{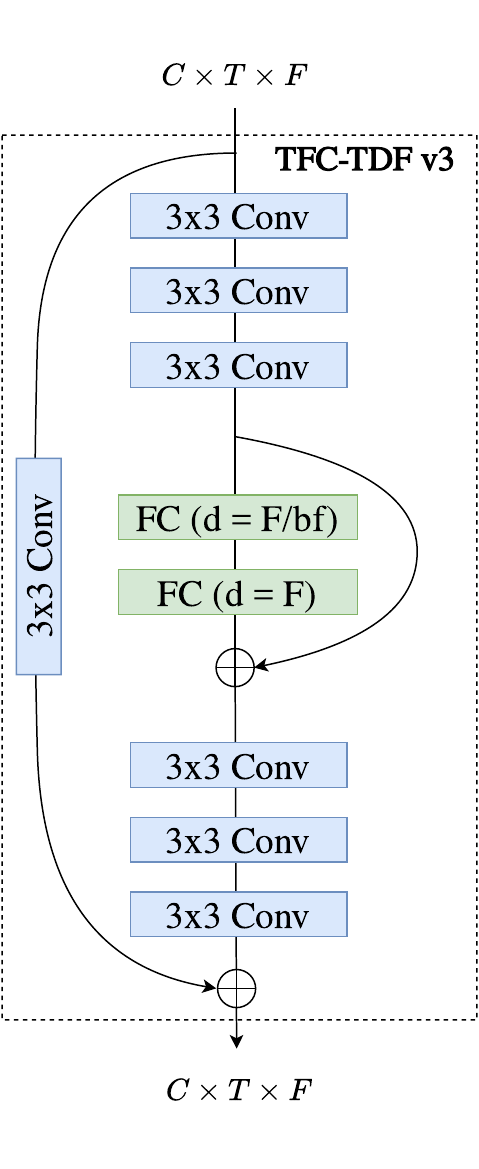}
        \label{fig:v3}
    }
    \hspace{0.1\linewidth}
    \subfigure[Improved Dual-Path Module]{
        \includegraphics[height=0.4\linewidth]{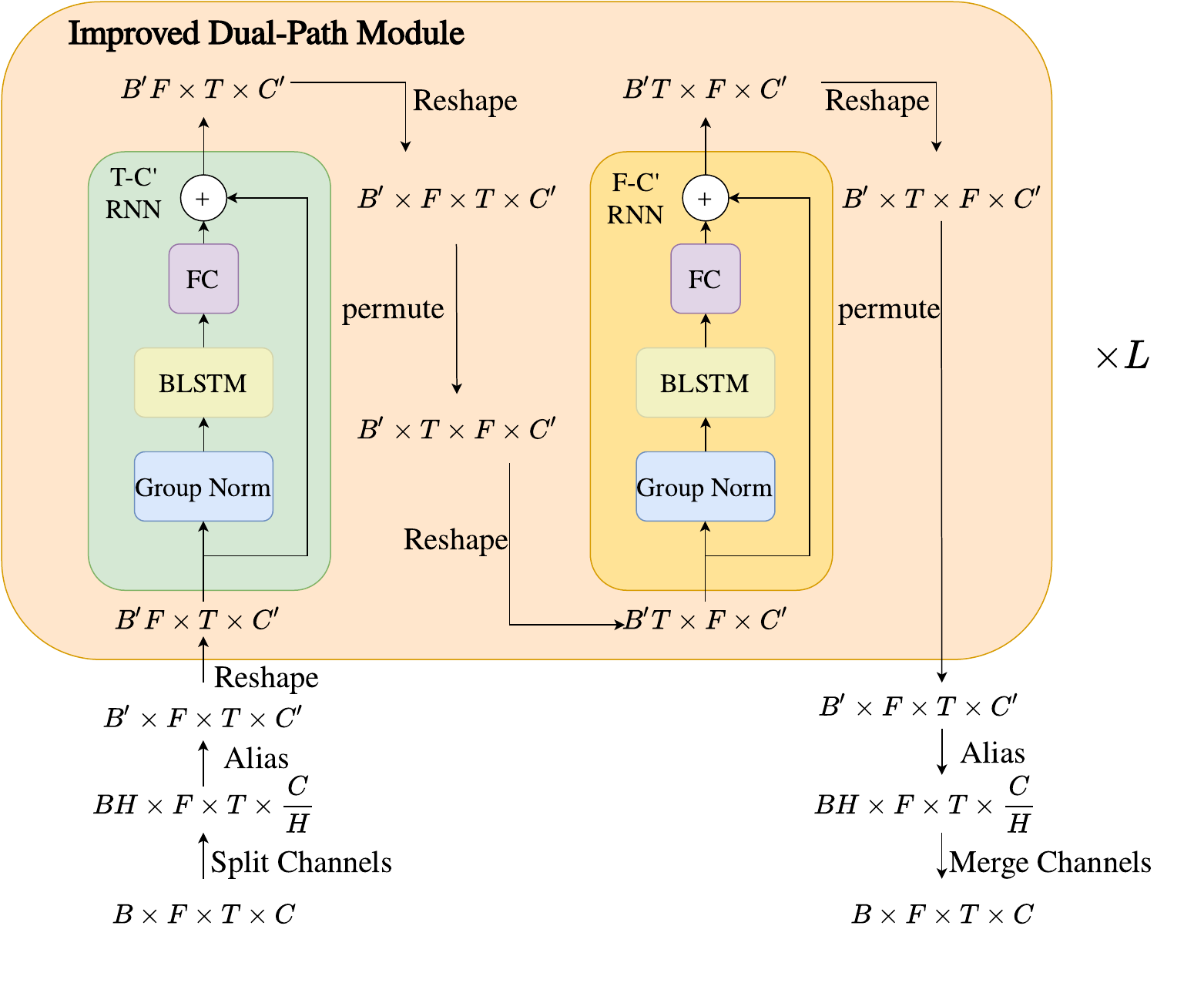}
        \label{fig:fsm}
    }
    \caption{Sub-blocks of DTTNet, where $bf$ is the bottleneck factor of Time Distributed Fully-connected layer (TDF); $B$ is the batch size; $ F$ is the number of features on the frequency axis; $T$ is the number of features on the time axis; $C$ is the number of channels generated by the convolution layer; 
    $L$ is the number of repeats of IDPM.}
    \label{fig:submods}
\end{figure*}

\begin{enumerate}[topsep=0pt,itemsep=-1ex,partopsep=1ex,parsep=1ex]
\item  As shown in Fig.~\ref{fig:arch}, by integrating and optimizing the encoder and decoder from TFC-TDF UNet v3 and the latent Dual-Path module from BSRNN, we reduce redundant parameters.

% We employ TFC-TDF v3 for both the UNet encoder and decoder as shown in Fig.~\ref{fig:v3}, 
\item As shown in Fig.~\ref{fig:fsm}, we partition the channels $C$ within the improved Dual-Path module, which reduces the inference time.

% We introduce the use of Dual-Path RNN to enhance the capture of inter-time and inter-frequency dependencies, simultaneously reducing model inference time through feature dimension splitting.
\item We optimize hyper-parameters within DTTNet and improve the Signal-to-Distortion Ratio (SDR) which is comparable to BSRNN~\cite{luo_music_2022} and TFC-TDF UNet v3~\cite{kim_sound_2023} as shown in Table~\ref{results}.
\item We test DTTNet with intricate audio patterns commonly misclassified by many models that are trained on MUSDB18-HQ dataset~\cite{MUSDB18}.
% Our model undergoes fine-tuning, focusing on  found in modern pop songs, followed by pattern-specific evaluations.
\end{enumerate}

\section{Dual Path TFC-TDF UNet}

% \subsection{Method Overview}
As depicted in Fig.~\ref{fig:arch}, our framework consists of 3 parts: encoder, decoder and latent part. The encoder and decoder are connected through skip connections which are identified as element-wise multiplications on the dotted arrows. For each layer inside the encoder and decoder, we use the residual convolution block (TFC-TDF v3)~\cite{kim_sound_2023}. The latent part is composed of a TFC-TDF v3 block and L layers of the Improved Dual-Path Module (IDPM). 

\subsection{Encoder}
The encoder initially uses 1x1 convolution block to increase the number of channels from $C$ to $g$. It is then, followed by $D$ layers, wherein each layer contains a TFC-TDF v3 block followed by a down-sampling block which consists of $3 \times 3$ convolution layer to reduce the feature map by half and increase the number of input channels by $g$. The inner structure of TFC-TDF v3 in each layer is shown in Fig.~\ref{fig:v3}. It has a Time-Frequency Convolutions (TFC) block, which contains 3 convolution blocks, followed by a residual Time-Distributed Fully-connected layer (TDF) where the frequency axis is first reduced by bottleneck factor $bf$ and then recovered to the original input dimension. This is followed by another TFC block. Finally, a residual connection from a single convolution layer is added to the output.

\subsection{Improved Dual-Path Module}
The Improved Dual-Path Module (IDPM) has a similar structure as the Band and Sequence Module in BSRNN~\cite{luo_music_2022}. This module is repeated $L$ times.

Fig.~\ref{fig:fsm} shows the structure of a single IDPM. To reduce the inference time while maintaining a high input dimension size $C$, we first split the input channel $C$ into $H$ heads. Then, $H$ heads are first processed along the time axis in the first RNN block and then along the frequency axis in the second RNN block. At the end of the IDPM, the reverse process is applied to merge the $H$ heads into $C$ channels.

In each RNN block, the split channels $C'$ are first normalized by a group normalization layer
(Group Norm)~\cite{wu_group_2018}, followed by a Bidirectional Long Short-Term Memory (BLSTM)~\cite{schuster_bidirectional_1997} which has input size $C'$ and hidden size $2C'$. The output size of BLSTM is $4C'$ and is reduced to $C'$ by a fully connected layer (FC). Finally, the residual is added to the output.

\subsection{Decoder}
Similar to the structure of the encoder, each layer in the decoder block contains an up-sampling block, which is a single $3 \times 3$ transposed convolution block that up-samples the feature map by a factor of two and decreases the number of input channels by $g$. At each layer, the up-sampled output is multiplied by the feature map from the encoder and the multiplied feature map is refined by a TFC-TDF v3 block without changing any of the shapes in the feature map. Finally, a 1x1 convolution block is applied to reduce the channels from $g$ to $C$.

\section{Generalization to Other Audio Patterns}
\newcommand{\picheight}{0.18\linewidth}
In this section, we explore the generalization ability of DTTNet trained on the 'vocals' of MUSDB18-HQ~\cite{MUSDB18} dataset for 5
intricate patterns, namely: \textit{Wah Guitar} (26 min), \textit{Horns} (1h 23 min), \textit{Sirens (2h 24 min)}, \textit{Up-filters} (37 min), and \textit{Vocal Chops} (42 min) taken from a bespoke dataset. The segments in each pattern are further divided into training (b-train), validation (b-valid), and test (b-test) sets in the ratios of 5:1:4. The sets consist of 4 to 8 second segments. The \textit{Vocal Chops} is included as the prediction target, whereas the other 4 patterns are not considered as targets for prediction\footnote{MUSDB18-HQ classifies \textit{Vocal Chops} as positive samples (e.g. 'PR - Happy Daze' in the test set). In the real-world scenario, they should be considered as accompaniment since pitch-tracking applications are only interested in human vocals.}.

For each training mixture chunk $m_t \in R^{C \times T}$ in MUSDB18-HQ, we randomly sample a segment $seg \in R^{C \times T'}$ from the b-train and the new mixture is defined as $m_t' = m_t + pad\_or\_truncate(seg)$. 

% The new target is defined as $t_t' = t_t + pad\_or\_truncate(seg)$ for \textit{Vocal Chops} and $t_t' = t_t$ for the other 4 patterns.

For each song $s_v \in R^{C \times T}$ in MUSDB18-HQ validation set, we randomly select a subset of segments $l_s \subset \{R^{C \times T'}\}$ from b-valid and create $s_v' = s_v + concat\_and\_pad(l_s)$ such that the 55\% of $concat\_and\_pad(l_s)$ are intermediately padded with zeros. A similar process is followed for the test set.
% A similar blending process is followed for b-valid and b-test. But we ensure that the total segments are intermediately zero-padded and the valid takes 

% they select and concatenate the segments to form a synthetic song. And the synthetic song has 55\% of intermediate zero padding to match the length of the original song. 
% During the training time, we select a pattern, from which we select a segment and blend it with the training chunk from MUSDB18-HQ. All selections are randomized. 

% We also randomly select and concatenate the segments to form a synthetic song for.  is mixed with the original song. And the synthetic song has 55\% of intermediate zero padding to match the length of the original song. Similarly, the test set is also formed in this manner.

% The segments from the validation set are randomly selected and concatenated to form a synthetic song of 45\% length of each song in the validation set from MUSDB18-HQ. Similarly, the test set is also formed in this manner.

% in a non-overlapped manner and roughly take 
Since \textit{Vocal Chops} are pitched and distorted, the untuned DTTNet (DTT) is fine-tuned when sampling from b-train in two ways. The first one takes the pattern \textit{Vocal Chops} into account (DTT + VC). The second one does not take \textit{Vocal Chops} into account (DTT + NVC).

% is fine-tuning DTTNet with \textit{Vocal Chops} (VC)  and the second one is fine-tuning with no \textit{Vocal Chops} (NVC). The Vocal chops are treated as positive samples\footnote{MUSDB18-HQ classifies \textit{Vocal Chops} as positive samples (e.g. 'PR - Happy Daze' in the test set). In the real-world scenario, they should be considered as accompaniment since pitch tracking applications are only interested in human vocals.} and the other 4 patterns are negative samples for the target track 'vocals'. Finally, we fine-tune our model along with the MUSDB18 training set.  

% \footnote{See song 'PR - Happy Daze' and 'PR - Oh No'}, we also add vocal chops into the prediction target\footnote{In the real-world scenario they should be considered as accompaniment since pitch tracking applications are only interested in human vocals}, while other samples are added into the mixture. Finally, we fine-tune our model along with the MUSDB18 training set.  

\section{Experiment}
\subsection{Dataset}

The MUSDB18-HQ dataset~\cite{MUSDB18} consists of 150 songs, each sampled at 44100 Hz with 2 channels. Each song contains 4 independent tracks: 'vocals', 'drums', 'bass', and 'other'. For our experiment, the dataset is split into a training set with 86 songs, a test set with 50 songs, and the remaining 14 songs are left for hyperparameter tuning. For data augmentation, we use pitch shift semitone in \{-2, -1, 0, 1, 2\} and time stretch with the percentage in \{-20, -10, 0, 10, 20\}.

\subsection{Experimental Setup}
We use AdamW optimizer with learning rate $2 \times 10^{-4}$ for our model. The model is trained on 6 second chunks with L1 loss rather than L2 on waveforms, since some of the training data is unclean\footnote{E.g. 'The So So Glos - Emergency' and 'The Districts - Vermont' at 1:44.}. We use two A40 GPUs with batch sizes of 8 each. For MUSDB18-HQ training, the epoch size is set to 3240 chunks and the model is trained for 4082 epochs. For fine-tuning, the epoch size is set to 324 chunks and the model is trained for 300 epochs. In both scenarios, we picked the model that has the best Utterance-level SDR (uSDR)~\cite{mitsufuji_music_2022} on the validation set. 

For STFT, we use a window size of 6144 and hop length 1024. For 'vocals', 'drums' and 'other', we cut the frequency bins to 2048~\cite{kimKUIELabMDXNetTwoStreamNeural2021}. We set the bottleneck factor $bf=8$ in TDF of Fig.~\ref{fig:v3}. For 'bass', we cut the frequency bins to 864 and we set $bf=2$ in TDF. For IDPM, we set L = 4 and ensure that each Group Norm has 16 channels~\cite{wu_group_2018}. We set $g=32$ for the up/down-sampling layers as shown in Fig.~\ref{fig:arch}.

\subsection{Evaluation Metrics}
We use Source-to-Distortion Ratio (SDR) for performance measurement.

\noindent \textbf{Chunk-level SDR (cSDR)}~\cite{SiSEC18} calculates SDR on 1 second chunks and reports the median value.
% We use it to compare with SOTA following the convention.

\noindent \textbf{Utterance-level SDR (uSDR)}~\cite{mitsufuji_music_2022} calculates the SDR for each song and reports its mean value.
% , which is more sensitive to the quality of the prediction result. We use it for validation and fine-tuning evaluation.

\section{Results and Discussion}
\subsection{Hyper-parameters and Performance}
In this section, we experiment with different combinations of hyper-parameters and study the impact on the uSDR and inference time. The results are presented in Table~\ref{tbl:hyer}.

\noindent \textbf{TFC-TDF Block} We noticed that TFC-TDF v3 residual block is more effective than TFC-TDF v2, significantly improving uSDR with minimal impact on training time.

\noindent \textbf{Heads and Channels} Increasing the number of channels $g$ from 32 to 48 offers slightly better performance but results in 4x more inference time. It was observed that inference time is appreciably reduced by setting $H=2$ compared to $H=1$ provided $g=32$. However, it takes double the epochs to attain the same level of uSDR while maintaining similar training times.

\noindent \textbf{Layers of IDPM} Increasing the repeats $L$ of IDPM significantly increases the inference time, however, the SDRs tend to decrease. Hence, setting $L = 4$ is sufficient to achieve good SDRs and inference time as compared to $L = 10$.
% as our feature encoder and decoder are stronger than the FC/MLP in BSRNN.
% In contrast, BSRNN relies on simpler FC and MLP methods, necessitating more layers $L$ and parameters.

\begin{table}[h]
    \centering
    \caption{Validation Set uSDR for 'vocals'. Experiments are carried out on mono audio (mid-channel) with 1200 epochs. The inference time is measured on a single RTX 3090 (24 GB) with a 3 minute input audio. Batch size is set to 4.}\label{tbl:hyer}
    \begin{tabular}{cccc|cc}\hline
        g& H &  L& block &uSDR & inference time\\\hline
        32 &2 &  4& v2 & 10.14 & \textbf{1.77} s\\
        % 32 &2 &  4& v2&9.52&9.68& - \\
        \textbf{48}& 2 & 4& v2  & 10.47 & 15.80 s\\
        32 &\textbf{1} & 4& v2 & 10.26 & 12.96 s\\
        32 & \textbf{1} &  \textbf{10}& v2  & 10.16 & 30.65 s\\
        32 &2 &  4& \textbf{v3}&\textbf{10.51}& 2.39 s
        \\\hline
    \end{tabular}
	
\end{table}

% \begin{table}[h]
%     \centering
%     \caption{Test Set SDR for vocal track, experiments are done on mono audio (mid-channel). The inference time is measured on a single RTX 3090 (24 GB) with a 3 minute input. Batch size is set to 4.}\label{tbl:hyer}
%     \begin{tabular}{cccc|ccc}\hline
%         g& H &  L& block&cSDR &uSDR & inference time\\\hline
%         32 &2 &  4& v2&9.24&9.47& 3.6 s\\
%         % 32 &2 &  4& v2&9.52&9.68& - \\
%         \textbf{48}& 2 & 4& v2&9.63&9.69& 15.8 s\\
%         32 &\textbf{1} & 4& v2&9.38&9.57& 12.9 s\\
%         32 & \textbf{1} &  \textbf{10}& v2&9.46&9.52& 30.6 s\\
%         32 &2 &  4& \textbf{v3}&\textbf{9.76}&\textbf{9.89}& \textbf{2.34} s
%         \\\hline
%     \end{tabular}
	
% \end{table}

\subsection{Study on Generalization Ability}
As depicted in Table~\ref{tbl:gen}, for untuned DTTNet, the uSDRs of the intricate test patterns in the bespoke dataset are lower compared to MUSDB18-HQ test set. The uSDR for \textit{All} patterns (MUSDB18-HQ + intricate patterns) is 5.64 dB lower than the MUSDB18-HQ for untuned DTTNet. 
% Notably, the 'sirens' pattern exhibited the most significant decrease in uSDR, primarily due to its dynamic nature.

For each pattern, DTT + VC outperforms DTT + NVC. On the MUSDB18-HQ test set, we observed 0.07 dB uSDR improvement for DTT + VC compared to DTT. When considering all the patterns, the uSDR for DTT + VC is 5.67 dB higher than the uSDR of DTT possibly because of the diversity of the patterns during training. Additionally, for \textit{Vocal Chops}, DTT + NVC shows lower uSDR possibly because this pattern does not appear in the training data leading to overfitting.

Furthermore, although we obtained 0.07 dB uSDR improvement on the MUSDB18-HQ test set using DTT + VC, cSDR drops by 0.12 dB compared to DTT. This suggests that our bespoke dataset is slightly smaller, posing a risk of overfitting.

% Fine-tuning model without vocal chops (NVC) slightly outperforms fine-tuning with vocal chops (VC) on most patterns, likely due to a higher sampling frequency (25\% vs. 20\%). However, the model struggles with vocal chops and shows poor results on vocal tracks. Additionally, it tends to overfit and lacks confidence when predicting complex patterns like chords. On the other hand, VC excels in NVC when combining all patterns because it encounters a wider variety of patterns during training.

% Furthermore, we observed a decrease in uSDR on the original test set following model fine-tuning. This suggests that our collected sample length is slightly insufficient, posing a risk of overfitting.

% \begin{figure}[h]
%     \centering
%     \includegraphics[width=0.9\linewidth]{pdfs/vocals_uSDR.pdf}
%     \caption{uSDR on various test sets.}
%     \label{fig:gen}
% \end{figure}

\begin{table}[h]
    \centering
    \caption{uSDR results on various test sets, with the best performance highlighted in boldface.}\label{tbl:gen}
    \begin{adjustbox}{width=0.45\textwidth}
    \begin{tabular}{cccc}\hline
         %   \multirow{2}{*}{Test Set}
         % &\multicolumn{3}{c}{uSDR}\\\cline{2-4}
           
        Test Set &DTT &DTT + NVC &DTT + VC\\\hline

MUSDB18-HQ & 10.15 & 10.07 & \textbf{10.22} \\ 
+ Wah Guitar & 7.97 & 9.73 & \textbf{10.04} \\ 
+ Horns & 8.23 & 9.95 & \textbf{10.14} \\ 
+ Sirens & 6.07 & 9.71 & \textbf{9.91} \\ 
+ Up-filters & 8.56 & 9.72 & \textbf{9.97} \\ 
+ Vocal Chops & 9.90 & 9.10 & \textbf{10.87} \\ 
+ All & 4.51 & 8.38 & \textbf{10.18} \\ 
\hline
    \end{tabular}
    \end{adjustbox}
	
\end{table}

% \vspace{-20pt}
\subsection{Comparison with the State-of-the-art (SOTA)}
As indicated in Table~\ref{results}, our DTTNet achieves higher cSDR on the 'vocals' track against BSRNN (SOTA) with only 13.3\% of its parameter size. Moreover, we also achieved a higher cSDR on the 'other' track compared to TFC-TDF UNet v3 (SOTA) with only 28.6\% of its parameter size.
\begin{table}[h]
    \centering
    \caption{cSDR in dB on MUSDB18-HQ Test Set. For parameter size (Params), we measure the single source model times the number of sources. $^\star$ uses bag of 4 models. The parameter of $^\dagger$ is measured based on the available code.\protect\footnotemark}\label{results}
    \begin{adjustbox}{width=0.48\textwidth}
    \begin{tabular}{c cccc c}\hline
        Model &vocals &drums & bass & other  & Params
        \\\hline

        % \textbf{x} & a & b & a & a& a& a& a& a& a M\\\hline
        ResUNetDecouple+~\cite{kong_decoupling_2021}& 
        8.98 & 6.62 & 6.04 & 5.29  & 102.0 M $\times$ 4\\
        
        CWS-PResUNet~\cite{liu_cws-presunet_2021} & 
        8.9 &  6.38  & 5.93 & 5.84  & 202.6 M $\times$ 4\\
        
        Hybrid Demucs (v3)$^\star$~\cite{defossez_hybrid_2021} & 
        8.13 &  8.24 &  \textbf{8.76} &5.59  & 83.6 M $\times$ 4\\
        
        HT Demucs~\cite{rouard_hybrid_2022} & 
        7.93 &  7.94 &  8.48 & 5.72  &  42.0 M\\
        
        BSRNN$^\dagger$~\cite{luo_music_2022} & 
        10.01 &  \textbf{9.01} &  7.22& 6.70& 37.6 M $\times$ 4\\
        
        KUIELab-MDX-Net~\cite{kimKUIELabMDXNetTwoStreamNeural2021} &8.97  &7.20  &7.83 &5.90  & 7.4 M $\times$ 4\\
        
        TFC-TDF UNet v3~\cite{kim_sound_2023} & 
        9.59 &  8.44 &  8.45 & 6.86 & 70.0 M\\\hline

        \textbf{DTTNet (Proposed)} & \textbf{10.12} & 7.74 & 7.45 & \textbf{6.92} & \textbf{5.0 M $\times$ 4} \\ 
        % DualPath TFC-TDF (L) & 9.95 & 7.55 & 7.35 & \textbf{6.87}  & 7.79 M $\times$ 4 \\
        \hline 
    \end{tabular}
    
    \end{adjustbox}
	
\end{table}
\footnotetext{\url{https://github.com/amanteur/BandSplitRNN-Pytorch}}

% \vspace{-20pt}
\section{Conclusion}

We introduce DTTNet, a novel and lightweight framework with higher cSDR but with reduced parameter size compared to both BSRNN and TFC-TDF UNet v3 for the 'vocals' and 'other' track music source separation respectively. Furthermore, we created a bespoke dataset of intricate patterns such as \textit{Vocal Chops} and tested the generalization ability of DTTNet.

In our future work, we plan to improve our framework to enhance the SDR for the 'drums' and 'bass' tracks. Additionally, we plan to integrate zero-shot systems~\cite{chen_zero-shot_2022} as a post-processing module to improve the generalization ability.

\section{Acknowlegement}
We acknowledge Dr. Lorenzo Picinali for his feedback on the initial background work, and Dr. Aidan O. T. Hogg and Shiliang Chen for writing and submission tips for ICASSP.

%

% \input{texs/others}

% \vfill\pagebreak

% \section{REFERENCES}
% \label{sec:refs}

% List and number all bibliographical references at the end of the
% paper. The references can be numbered in alphabetic order or in
% order of appearance in the document. When referring to them in
% the text, type the corresponding reference number in square
% brackets as shown at the end of this sentence \cite{C2}. An
% additional final page (the fifth page, in most cases) is
% allowed, but must contain only references to the prior
% literature.

\bibliographystyle{IEEEbib}
\bibliography{strings,refs}

\end{document}